# Discrete scale invariance of the quasi-bound states at atomic vacancies in a topological material

Zhibin Shao[a,1], Shaojian Li[b,1], Yanzhao Liu[c,1], Zi Li[d,1], Huichao Wang[e], Qi Bian[b], Jiaqiang Yan[f], David Mandrus[f,g], Haiwen Liu[h], Ping Zhang[d,i,2], X.C. Xie[c,j,k,l], Jian Wang[c,j,k,l,2] and Minghu Pan[a,b,2]

[a]School of Physics and Information Technology, Shaanxi Normal University, Xi'an 710119, China; [b]School of Physics, Huazhong University of Science and Technology, Wuhan 430074, China; [c]International Center for Quantum Materials, School of Physics, Peking University, Beijing 100871, China; [d]Institute of Applied Physics and Computational Mathematics, Beijing 100088, China; [e]School of Physics, Sun Yat-sen University, Guangzhou 510275, China; [f]Materials Science and Technology Division, Oak Ridge National Laboratory, Oak Ridge, TN 37831, USA; [g]Department of Materials Science and Engineering, University of Tennessee, Knoxville, TN 37996, USA; [h]Center for Advanced Quantum Studies, Department of Physics, Beijing Normal University, Beijing 100875, China; [i]School of Physics and Physical Engineering, Qufu Normal University, Qufu 273165, China; [j]Collaborative Innovation Center of Quantum Matter, Beijing 100871, China; [k]CAS Center for Excellence in Topological Quantum Computation, University of Chinese Academy of Sciences, Beijing 100190, China; [l]Beijing Academy of Quantum Information Sciences, Beijing 100193, China.

[1]Z.S., S.L., Y.L. and Z.L. contributed equally to this work.

[2]To whom correspondence may be addressed. Email: jianwangphysics@pku.edu.cn (Jian Wang), minghupan@snnu.edu.cn (Minghu Pan), zhang_ping@iapcm.ac.cn (Ping Zhang)

## Abstract

**Recently, log-periodic quantum oscillations have been detected in topological materials zirconium pentatelluride ($ZrTe_5$) and hafnium pentatelluride ($HfTe_5$), displaying intriguing discrete scale invariance (DSI) characteristic. In condensed materials, the DSI is considered to be related to the quasi-bound states formed by massless Dirac fermions with strong Coulomb attraction, offering a feasible platform to study the long-pursued atomic-collapse phenomenon. Here, we demonstrate that a variety of atomic vacancies in the topological material $HfTe_5$ can host the geometric quasi-bound states with DSI feature, resembling the artificial supercritical atom collapse. The density of states of these quasi-bound states are enhanced and the quasi-bound states are spatially distributed in the "orbitals" surrounding the vacancy sites, which are detected and visualized by low-temperature scanning tunneling microscope/spectroscopy (STM/S). By applying the perpendicular magnetic fields, the quasi-bound states at lower energies**

become wider and eventually invisible, meanwhile the energies of quasi-bound states move gradually towards the Fermi energy ($E_F$). These features are consistent with the theoretical prediction of a magnetic-field-induced transition from supercritical to subcritical states. The direct observation of geometric quasi-bound states sheds light on the deep understanding of the DSI in quantum materials.

## Significance Statement

The atomic collapse (AC) state in superheavy atoms is one of the most fundamental problems remaining elusive in experiments. In materials with relativistic Dirac carriers, quasi-bound states featuring discrete scale invariance are proposed as the analogue of the AC state. Here, by using scanning tunneling microscope/spectroscopy, we detect a series of differential conductance peaks at geometric energies on atomic vacancies in topological material hafnium pentatelluride ($HfTe_5$), which represent discrete scale invariant quasi-bound states. The geometric spatial distribution of these states is also detected. Our experimental observations are consistent with the AC model in Dirac materials. The direct imaging of geometric quasi-bound states at the atomic scale sheds light on the deep understanding of the scale anomaly and AC state in quantum materials.

## Introduction

One of the most striking and challenging issues in fundamental physics is the observation of supercriticality phenomenon in ultra-heavy nuclei which does not exist in nature. The atomic wave function will collapse when the nuclear charge parameter $Z$ exceeds a critical value $Z_c$ (1-4). Unfortunately, the very large $Z_c$ makes it almost impossible for atoms to satisfy the supercritical condition that $\beta=Z\cdot\alpha$ exceeds a critical value of order unity, where $\alpha\sim1/137$ is the fine structure constant. Recently, the Dirac materials with massless or massive Dirac fermions analogous to high energy relativistic particles provide new platforms for related investigations. Moreover, the supercritical regime guarantees the existence of a geometric series of quasi-bound states showing dramatic discrete scale invariance (DSI) property (5,6). The DSI induces log-periodic corrections to scaling and has been rarely demonstrated in quantum systems other than cold atom gas (7-14). In condensed materials, the atomic collapse states were observed in graphene by scanning tunneling microscope/spectroscopy (STM/S) experiments (15,16) and further STM/S studies of graphene reported the two quasi-bound states indicating a signature of DSI (17). In addition, the DSI feature has been clearly detected in topological materials zirconium pentatelluride ($ZrTe_5$) and hafnium pentatelluride ($HfTe_5$) by the observation of log-periodic quantum oscillation involving up to five oscillating cycles in the magnetoresistance (MR) under ultrahigh magnetic fields ($B$) (18-20). The underlying mechanism can be attributed to the discrete scale invariant quasi-bound states composed of relativistic quasiparticles and non-relativistic quasiparticles (or charge impurity) in Dirac materials (18,21). Thus, Dirac materials can serve as promising systems to uncover the DSI characteristic in quantum systems that satisfy the supercritical collapse condition. Direct imaging of the quasi-bound states responsible for the appearance of the DSI feature in topological systems is certainly of particular importance while it has not been presented.

$ZrTe_5$ and $HfTe_5$ are predicted to be quantum spin Hall insulators in the two-dimensional (2D) limit and the three-dimensional (3D) crystals are located near the phase boundary between weak and strong topological insulators (TIs) (22,23). However, previous experimental results about these materials present a large diversity. Some angle-resolved photoemission spectroscopy (ARPES) (24,25) and magnetoinfrared spectroscopy studies show that $ZrTe_5$ is a Dirac semimetal (26). Negative magnetoresistance and the anomalous Hall effect also support the massless Dirac band structure of $ZrTe_5$ (27-29). Recently, more and more electrical transport, ARPES and STM results suggest that $ZrTe_5$ and $HfTe_5$ are TIs with a small gap (30-35). The small Fermi surface and ultralow carrier density of $ZrTe_5$ and $HfTe_5$ guarantee the supercritical Coulomb attraction due to the weak screening effect (18), which makes them promising platforms to explore the quasi-bound states with DSI.

## Results

**Atomic-resolved STM image and small energy gap of HfTe$_5$ surface.** The HfTe$_5$ crystal shows an orthorhombic layered structure (36) and contains 2D sheets of HfTe$_5$ in the *a-c* plane, which stack along the *b* axis *via* interlayer *van der Waals* interactions, as shown in Fig. 1*A*. Each 2D sheet consists of alternating prismatic HfTe$_3$ chains along *a* axis that are linked by parallel zigzag Te chains. The prism of HfTe$_3$ is formed by a dimer of Te and an apical Te atom surrounding a Hf atom. HfTe$_5$ is easily cleavable along the *a-c* plane, and exhibits a quasi-1D preference along *a* axis. Fig. 1*B* shows the atomically resolved topography of such cleaved HfTe$_5$ surface, in which 1D chains of Te dimers can be well-identified. The lattice constants along *a* and *c* axis can be directly measured from STM images, which are 3.93 and 13.5 Å, respectively, in good agreement with the crystal structure of HfTe$_5$ (22). The local density of states (LDOS, from spectroscopic measurement) obtained at the surface terrace are plotted in Fig. 1*C* at 77 K and Fig. 1*D* at 4.2 K. The d*I*/d*V* spectrum at 77 K shows a small energy gap developed around the Fermi energy ($E_F$). Further *I*/*V* and d*I*/d*V* spectroscopic measurements are performed at 4.2 K and repeated over forty times at different surface locations, as shown in Fig. 1*D*. As we see, an energy gap with the size of about 16 meV is identified with the top of the valence band and the bottom of the conductance band located at $\sim$ -1 and $\sim$ +15 meV, respectively. The observation of the TI gap in HfTe$_5$ is generally consistent with previous STM/ARPES work (37,38). The size of the gap observed on the HfTe$_5$ surface agrees well with the density-functional theory calculation on our Te-flux grown HfTe$_5$ samples (30). Note that the reported gap values by STS measurement in Ref. (31,33) on the ZrTe$_5$ surface are much larger than the value of HfTe$_5$ we reported here, which could be due to the difference of the samples.

**Surface defects and the calculated charges.** On the cleaved (010) surface of HfTe$_5$, there are a number of surface defects. These surface defects provide natural, static charge centers for generating the long-range Coulomb attraction considering the low carrier density in HfTe$_5$ (19). In our HfTe$_5$ sample, a very small gap around $E_F$ is detected (Fig. 1$D$), which indicates that our samples are TIs, consistent with previous photoemission reports (37). Theoretically, the small gap has little influence on the formation of quasi-bound states showing the DSI feature (19), and the quasi-bound states can be attributed to the Dirac fermions from the bulk massive Dirac bands (Fig. 1$C$, *Inset*). In this situation, the Weyl equation with Coulomb attraction $V(\vec{R}) = \frac{-Ze^2}{4\pi\varepsilon R}$ may obey the DSI property (18). Here $Ze$ is the central charge, and the fine structure constant $\alpha = \frac{e^2}{4\pi\varepsilon\hbar v_F}$ is larger than one due to the small Fermi velocity $v_F$ in Dirac materials compared with the speed of light in the vacuum (18,24,26,30,39-41). Previous transport studies have reported that the fine structure constant α is about 4.9 in our HfTe$_5$ samples (19). The so-called supercritical condition ($Z \cdot \alpha$ surpassing the angular momenta $m$) can be readily matched in this material and further gives rise to quasi-bound states solution in the system (5,6). If only considering the lowest angular momentum channel with $m = 1$, the charge number $Z$ of defects has to be larger than 0.20, in order to result in the quasi-bound states with DSI. Such DSI property of the quasi-bound states composed of charge defects and hole carriers in our HfTe$_5$ is supposed to give rise to log-periodic DOS peaks above $E_F$, which can be detected by differential tunneling spectroscopy, namely d$I$/d$V$ spectrum.

The variety of surface defects could provide a fertile playground to study the supercritical ($Z \cdot \alpha > m$) and subcritical ($Z \cdot \alpha < m$) phenomena in this intriguing material. To understand the defects-induced DSI effects, we systematically investigate the surface defects observed in experiments, rationalized by employing density-functional theory (DFT) calculations to estimate the local charge of the defect. As shown in Fig. 2$A$, the pristine HfTe$_5$ is first calculated. HfTe$_5$ has two kinds of Te atoms. Three Te atoms are in HfTe$_3$ prism, two Te are connecting in between two HfTe$_3$ prisms. We name the top Te atoms in the HfTe$_3$ prism, bottom Te atom in HfTe$_3$ prism and Te atoms connecting two adjacent HfTe$_3$ prisms as $Te_t$, $Te_d$ and $Te_z$, respectively. For the pristine surface, the HfTe$_5$ cell is neutral and the net charge is zero. Then we calculate the structures, simulated images and the charges for three different surface vacancies experimentally observed with a number of theoretically predicted defect structures, as shown in Fig. 2$B$-2$D$. For defect I, a single $Te_t$ vacancy could induce a net charge of 0.16 e (Fig. 2$B$). The surface $Te_t$ vacancies are carefully introduced to best fit the experimental STM image. More theoretically examined structures are listed in *SI Appendix,* Table S1. While for defects with larger net charges, the Hf vacancy should be included. Especially for defect II with the largest net charge, 2 Hf vacancies are introduced (Fig. 2*C*). To validate the proposed structures of these defects, the simulated STM images are also calculated, presented with the experimental STM image for each defect. The simulated STM images of the pristine surface and three surface vacancies are in nice consistency with the experiments.

**The observation of DSI quasi-bound states at atomic vacancies.** Fig. 3 summarizes high-resolution STM images and corresponding STS spectra for the pristine surface without defects (*A*) and three different surface defects (vacancies 1-3) with possible configurations, *e.g.* vac. 1, single *Te$_t$* vacancy (*B*), vac. 2, *2Hf+1Te$_d$* vacancy (*C*) and vac. 3, *1Hf* vacancy (*D*), respectively. For the pristine surface (Fig. 3*A* and *SI Appendix*, Fig. S1A) and *1Te$_t$* vacancy (Fig. 3*B* and *SI Appendix*, Fig. S1B), d*I*/d*V* spectra taken at 4.2 K show the clean curve without any extra peaks above $E_F$ in both linear and log scaled plots. Clearly, the local charge of *1Te$_t$* vacancy (*Z*) is only 0.165 electrons, which does not meet the supercritical condition $Z \geq 0.20$ electrons in HfTe$_5$. The charges of vac. 2 and vac. 3 estimated about 0.972 and 0.776, respectively, are much larger than the supercritical condition. Indeed, the d*I*/d*V* spectra of these defects show a series of DOS peaks above $E_F$ (Fig. 3*C*-3*D*). The differences of the height of resonance peaks and the overall background between Fig. 3$C$ (vac. 2) and Fig. 3$D$ (vac. 3) may be attributed to the local inhomogeneity and distinct kinds of vacancies as well as the charges. Theoretically, the energy of the quasi-bound states satisfies DSI when the Dirac point is chosen as the zero point (18). However, in our HfTe$_5$ sample, a very small gap around $E_F$ is detected. Consequently, the Dirac point is absent in the STS and the energy of the Dirac point should be replaced by the middle position of the gap $E_M$ (about 8 mV above the Fermi energy), which is very close to $E_F$. The log-scaled d$I$/d$V$ spectra as a function of $E$-$E_M$ are plotted in Fig. 3$C$-3$D$, which show clear log-periodicity and DSI of the quasi-bound states. The ratio $\lambda$ of $E_{n+1}$ $vs.$ $E_n$ is $\lambda = e^{\pi/s_0}$ (18), where $s_0 = \sqrt{(Z \cdot \alpha)^2 - 1}$. At around 10 mV, the resonance peaks slightly deviate from the log-periodicity, which may be induced by the influence of the energy gap. Based on the data at vac. 2 ($E$-$E_M$>30 mV) as shown in Fig. 3$C$, the scale factor $\lambda = 1.9$ gives the effective charge $Z \sim 1$ of vac. 2, which is consistent with the charge value (-0.972) from $ab$ $initio$ calculation. Furthermore, the value of $\lambda$ is consistent with the previous magnetoresistance measurements on our HfTe$_5$ samples (19), wherein the scale factor of the characteristic magnetic fields $\lambda_B = \frac{B_{n+1}}{B_n}$ is about 3.7 and then $\lambda_E = e^{\pi/s_0} = \sqrt{\lambda_B} \sim 1.92$. The $\lambda$ value of the DSI spectrum at vac. 3 shown in Fig. 3$D$ gives an estimation of charge value of about 1, which agrees roughly with the calculated charge -0.776. Note that the calculated net charge is slightly lower than the expected values, which may be due to the limited size of the simulation model and the definition of the defect charge.

In previous works on graphene, highly-charged impurities, *e.g.* calcium dimers created by tip manipulation (15), single atomic vacancy by applying voltage pulses or helium ion beam sputtering (16,17), serve as the superheavy atomic nucleus and two atomic collapse states were measured by tunneling spectroscopy. More atomic collapse states are still desired to make the DSI feature unambiguous. In crystalline topological material HfTe$_5$, intriguingly, we detected four remarkable quasi-bound states by STS above 19 meV (outside the small band gap ~ 16 meV) in different vacancies with the similar charge showing almost same $\lambda$ values. Furthermore, in another HfTe$_5$ sample, d$I$/d$V$ peaks at log-periodic energies are also detected at an atomic vacancy (vac. 4) with an effective charge $Z$ ~ 0.62, which is lower than $Z$ ~ 1 at vac. 2 and vac. 3 (*SI Appendix*, Fig. S2). The discrete scale invariant quasi-bound states at vacancies with different charges in different samples further support the emergence of geometric quasi-bound states at the atomic scale in HfTe$_5$.

**Spatial and magnetic field dependence of the quasi-bound states.** To further characterize the spatial dependence of the quasi-bound states that appear for the artificial nuclei, we performed d$I$/d$V$ mapping at a series of resonance energies around vac. 2 (Fig. 4$A$), *e.g.* +21, +41, +76 and +136 mV. The defects are randomly distributed and we find that vac. 2 is relatively far away from others. As shown in the top panel of Fig. 4$A$, the vacancy locates in the middle and there is no other defect along $a$ axis in the field of view. Furthermore, the intensity of the resonance extended outward from the vacancy center to a distance greater than 6 nm with decreasing the energies. The DOS is distributed mainly on the left and right sides of the vacancy, which may be due to the anisotropy of the Fermi velocity in the $a$-$c$ plane (40). Besides, d$I$/d$V$ maps at +76 and +136 mV display more symmetric DOS distribution around the vacancy than d$I$/d$V$ maps at +21 and +41 mV. The asymmetric distribution could be due to the distant defects (*SI Appendix,* Fig. S3), especially for the lower energy quasi-bound states (*e.g.* +21 mV) showing larger radii. We draw out the intensity of each quasi-bound state from the d$I$/d$V$ maps and plot them with lateral distances in Fig. 4$B$, which can be fitted with the radius $R_n$ (the distance from the center of the vacancy) for each quasi-bound state (See *SI Appendix* for more details). Due to the influence from other defects in the vicinity area, the uncertainties of radii at lower energies (+21 and +41 mV) are relatively large, and a precisely log-periodic relation is hard to determine. However, the spatial distribution of quasi-bound states still shows a log-periodic trend, especially for the quasi-bound states at high energies (+76 and +136 mV). The radii of quasi-bound states at +76 and +136 mV are estimated to be around 25 and 15 Å, respectively. The ratio between these two radii is about 1.7, which is close to the energy ratio of differential conductance peaks ($\lambda_E = \frac{E_{n+1}}{E_n} \sim 1.9$). In addition, if the dielectric constant of HfTe$_5$ $\varepsilon$ is of order 10 $\varepsilon_0$ (42), where $\varepsilon_0$ is the vacuum dielectric constant, the deduced $R_n$ at +76 and +136 mV from the relation $R_n \sim \frac{Ze^2}{4\pi\varepsilon E_n}$ (18) are 23 and 12 Å, which agree with the experimental results. These spatial properties of quasi-bound states further support the realization of quasi-bound states with DSI.

The evolution of d$I$/d$V$ spectra under magnetic fields is further investigated. Theoretically, the quasi-bound states with DSI are predicted to be converted into subcritical states with increasing magnetic fields (43) (see *SI Appendix* for more details). During the transition, the energy of the quasi-bound state gradually approaches the Fermi energy and the corresponding resonance peak becomes wider (*SI Appendix,* Fig. S4). The experimental results are shown in Fig. 4$D$-4$F$. The d$I$/d$V$ curve of vac. 3 presents four quasi-bound states at 0 T, which are determined by the Gaussian peak fitting (*SI Appendix,* Fig. S5A) and labeled as $E_1$, $E_2$, $E_3$ and $E_4$. As the magnetic field is increased, the resonance peak of the first quasi-bound state ($n$=1) becomes wider at 1 T with a larger error bar and is hard to be distinguished at 2 T (*SI Appendix,* Fig. S5B-S5C), in accordance with the magnetic-field-induced broadening effect in the transition. Similar magnetic-field-induced broadening effect is also observed for the second quasi-bound state ($n$=2) when the applied magnetic field is increased from 0 to 2 T, and the second quasi-bound state is almost indistinguishable at 3 T (*SI Appendix,* Fig. S5D). Note that, when the magnetic field is increased from 1 T to 2 T, the energy of the second quasi-bound state shown in Fig. 4$E$-4$F$ becomes smaller (approaching the Fermi energy) and deviates the log-periodicity (dashed line in Fig. 4*F*), which is also consistent with the magnetic-field-induced transition scenario (43). Furthermore, the resonance peak broadening and the energy shift to the Fermi energy with increasing magnetic fields are reproduced at vac. 2, as shown in *SI Appendix, Fig. S6-S7*.

When the magnetic field $B \geq 3$ T, only two quasi-bound states at higher energy (> 60 meV) are distinguishable, as shown in Fig. 4*D*-4*F*. The scale factor $\lambda$ of vac. 3 is about $\lambda = E_n/E_{n-1} \sim 2.0$. Therefore, the ratio between the characteristic fields $B_n/B_{n-1}$ should be $\lambda^2 \sim 4$ (19). Considering that the second quasi-bound state is indistinguishable at 3 T, it is expected that the resonance peak of the third quasi-bound state should disappear at $B > 12$ T, which exceeds the maximum magnetic field of our facilities.

## Conclusion

In conclusion, in quantum materials, two quasi-bound states in graphene were revealed by STM/S indicating the existence of DSI and transport measurements of topological materials showed a clear feature of DSI by detecting the log-periodic quantum oscillations in MR. The underlying quasi-bound states showing DSI call for more direct experimental evidence. Here, the STS results clearly demonstrate the geometric quasi-bound states around a variety of atomic vacancies in topological material HfTe$_5$. Four geometric quasi-bound states in different charge vacancies constitute the unambiguous evidence and a clean example for the realization of quasi-bound states with DSI in condensed matter. This work provides a promising platform for studying the supercritical atomic collapse with the relativistic effect in quantum systems.

## Materials and Methods

**Theoretical model and method.** To explore the possible atomic structures of the charged defects, we apply first-principles simulations to optimize the defect structures and calculate the defect charges. The simulations are performed using the plane-wave method, with the projector augmented wave pseudopotentials (44) and the Perde-Burke-Ernzerhof exchange-correlation functional (45) as implemented in the Vienna Ab initio Simulation Package (VASP) code (46,47). In the calculations, only the gamma point is sampled in the Brillouin zone, and the energy cutoff of the plane-wave basis is 300 eV.

To explore the defect structures, the simulation model is constructed as a single HfTe$_5$ layer in the $a$-$c$ plane, containing 6×4 unit cells (total of 144 atoms), and each unit cell includes 1 Hf atom and 5 Te atoms. The periodic boundary conditions are considered in the $a$ and $c$ directions, while in the $b$ direction, a 20 Å vacuum layer is added. The defect structures are designed as a complex of atomic vacancies with up to 2 Hf vacancies and 3 Te vacancies. The charge of the defect is calculated by summing the charges of all atoms in one or two HfTe$_5$ unit cells after removing the certain atoms; the atomic charge is obtained by Bader analysis (48) on the charge density obtained by the first-principles method. For the comparison with experiments, the STM images of the defects are also simulated theoretically, which are obtained by the profile of the partial charge density contributed by relevant electronic states (within 1.0 eV above the Fermi energy).

**Sample information.** Single crystals $HfTe_5$ used in our experiments were grown by the self-Te-flux method as described in the previous report (30). The samples in this work are of the same batch as that in Ref. (19), (20) and (35). Typical sizes of the $HfTe_5$ samples are about $15\,\mathrm{mm} \times 0.3\,\mathrm{mm} \times 0.1\,\mathrm{mm}$ (length × width × height). The longest direction of the sample is always along the crystallographic *a* axis and a large *a-c* plane can be obtained by exfoliation. The structure and composition of the $HfTe_5$ crystals were analyzed by powder X-ray diffraction, transmission electron microscopy and scanning electron microscopy with energy-dispersive X-ray spectroscopy (19,30,35). The carrier density and the mobility of the $HfTe_5$ samples at 2 K are estimated to be $1.4 \times 10^{15}\,\mathrm{cm}^{-3}$ and $1.84 \times 10^6$ cm$^2$V$^{-1}$s$^{-1}$ by electrical transport measurements, respectively (30).

**STM Characterization**. $HfTe_5$ are layered materials with *van der Waals* interlayer coupling (22). Therefore, a large atomically flat surface is conveniently obtained by exfoliation, which is suitable for STM measurements. Our STM experiments were carried out on an ultrahigh vacuum (UHV) commercial STM system (Unisoku) which can reach a temperature of 400 mK by using a single-shot $^3$He cryostat. The base pressure for experiments was $2.0 \times 10^{-10}$ Torr. HfTe$_5$ samples were cleaved in situ at 78 K and then transferred into STM. The bias voltage was applied on the samples. The STS data were obtained by a standard lock-in method that applied an additional small AC voltage with a frequency of 973.0 Hz. The d$I$/d$V$ spectra were collected by disrupting the feedback loop and sweeping the DC bias voltage. WSxM software was used for the post process of all STM data (49).

**Data Availability**. All study data are included in the article and/or *SI Appendix*.

**ACKNOWLEDGMENTS.** STM work was conducted at School of Physics, Huazhong University of Science and Technology in Wuhan. The calculation of the charges and structures of defects was conducted at Institute of Applied Physics and Computational Mathematics, Beijing. This work is financially supported by the National Key R&D Program of China (2018YFA0305604), the Postdoctoral Innovative Talent Support Program of China (No. BX20200202), the National Natural Science Foundation of China (11574095, 91745115, 11888101, 12004441, 92165204, 11625415, 12004234), the Beijing Natural Science Foundation (Z180010), the Strategic Priority Research Program of Chinese Academy of Sciences (XDB28000000), the Hundreds of Talents program of Sun Yat-Sen University, Fundamental Research Funds for the Central Universities (No. 202lqntd27) and the China Postdoctoral Science Foundation (2021M700253).

**Figures**

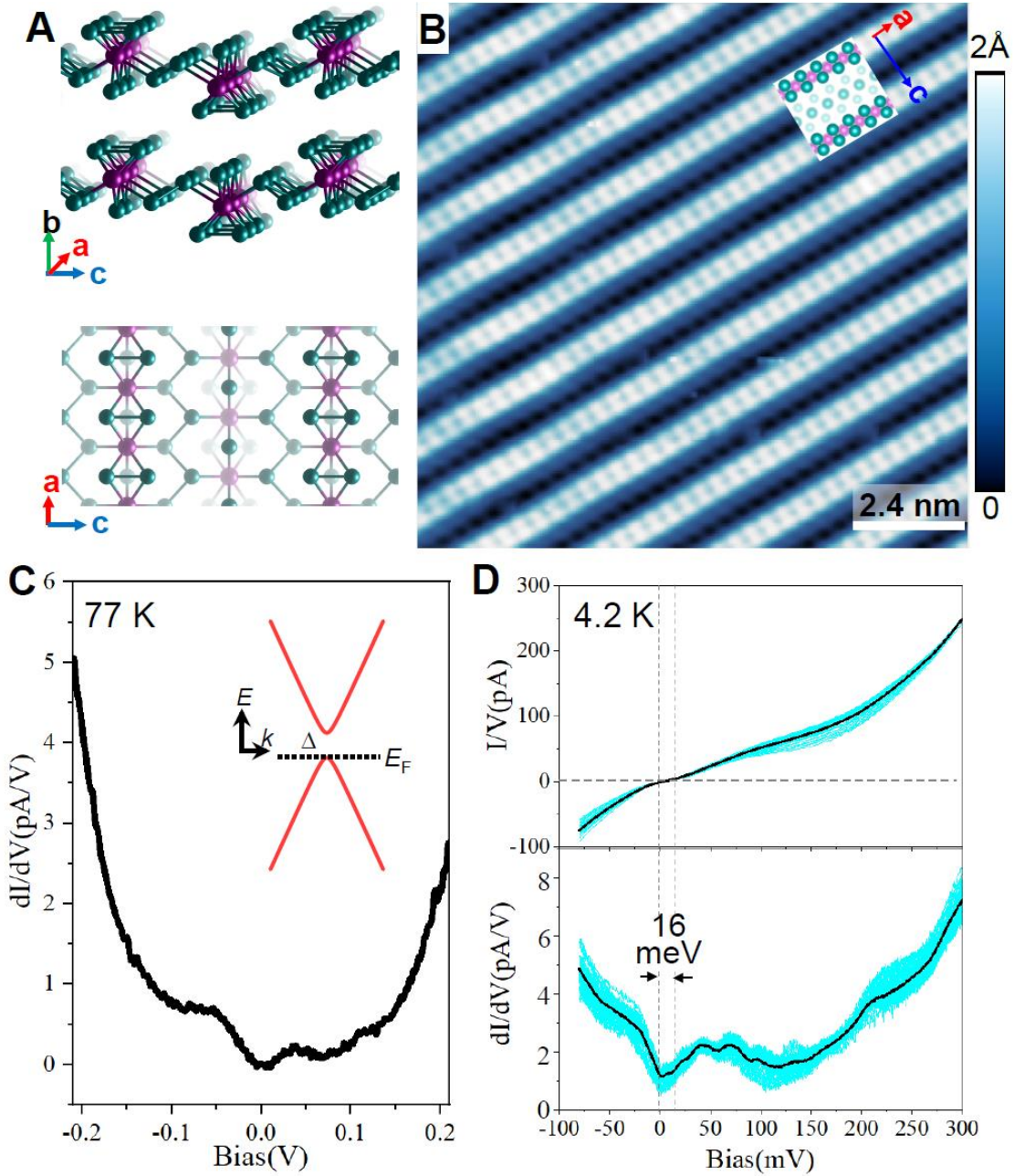

**Fig. 1. Structure, atomic-resolved topographic image and energy gap.**
(*A*) 3D structure of HfTe$_5$ crystal (upper) and top view of the single-layer structure (lower). (*B*) STM constant current topographic image of the cleaved *a-c* (010) surface of the HfTe$_5$ crystal ($I_T$ = 100 pA, $V_B$ = −800 mV). The image size is 12 nm × 12 nm. (*C*) Tunneling differential conductance spectrum taken at 77 K on the *a-c* surface ($I_T$ = 500 pA, $V_B$ = 500 mV) showing a small energy gap (Δ) around $E_F$. (*Inset*) Schematic of the HfTe$_5$ band structure with the gapped Dirac point. (*D*) $I/V$ conductance (upper) and d$I$/d$V$ differential conductance spectra measured at 4.2 K on the surface ($I_T$ = 300 pA, $V_B$ = - 300 mV). Such measurements were performed over 40 times at different surface locations. The cyan colored curves represent forty $I/V$ and d$I$/d$V$ spectra, while the averaged spectra are shown as the black curves.

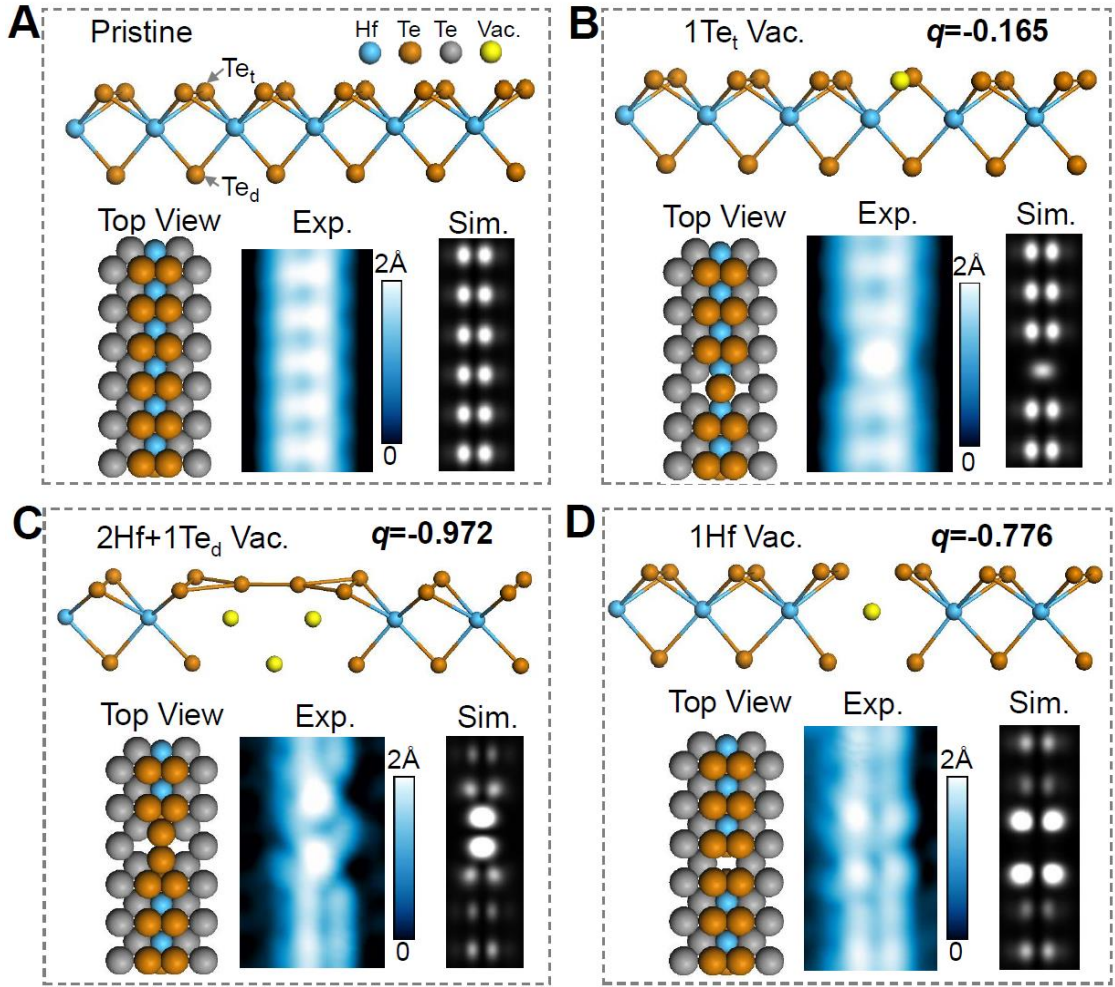

**Fig. 2. Structures, the charges and simulated images for pristine HfTe$_5$ and three different surface vacancies.** (*A*) Pristine HfTe$_5$ marked with two types of Te atoms with different coordination, (top) Te atoms in HfTe$_3$ prism (*Te$_t$*) and (bottom) Te atom in HfTe$_3$ prism (*Te$_d$*). (*B-D*) Structures, experimental / simulated STM images and the calculated charges for three theoretically predicted surface defects. For the sideview of the structures (the upper parts of panels *A-D*), only the HfTe$_3$ chain containing the vacancy is shown; the vacant atoms are labeled as yellow balls. For the top view, the Te atoms in the HfTe$_3$ chain and the zigzag Te chain are labeled as brown and gray balls, respectively.

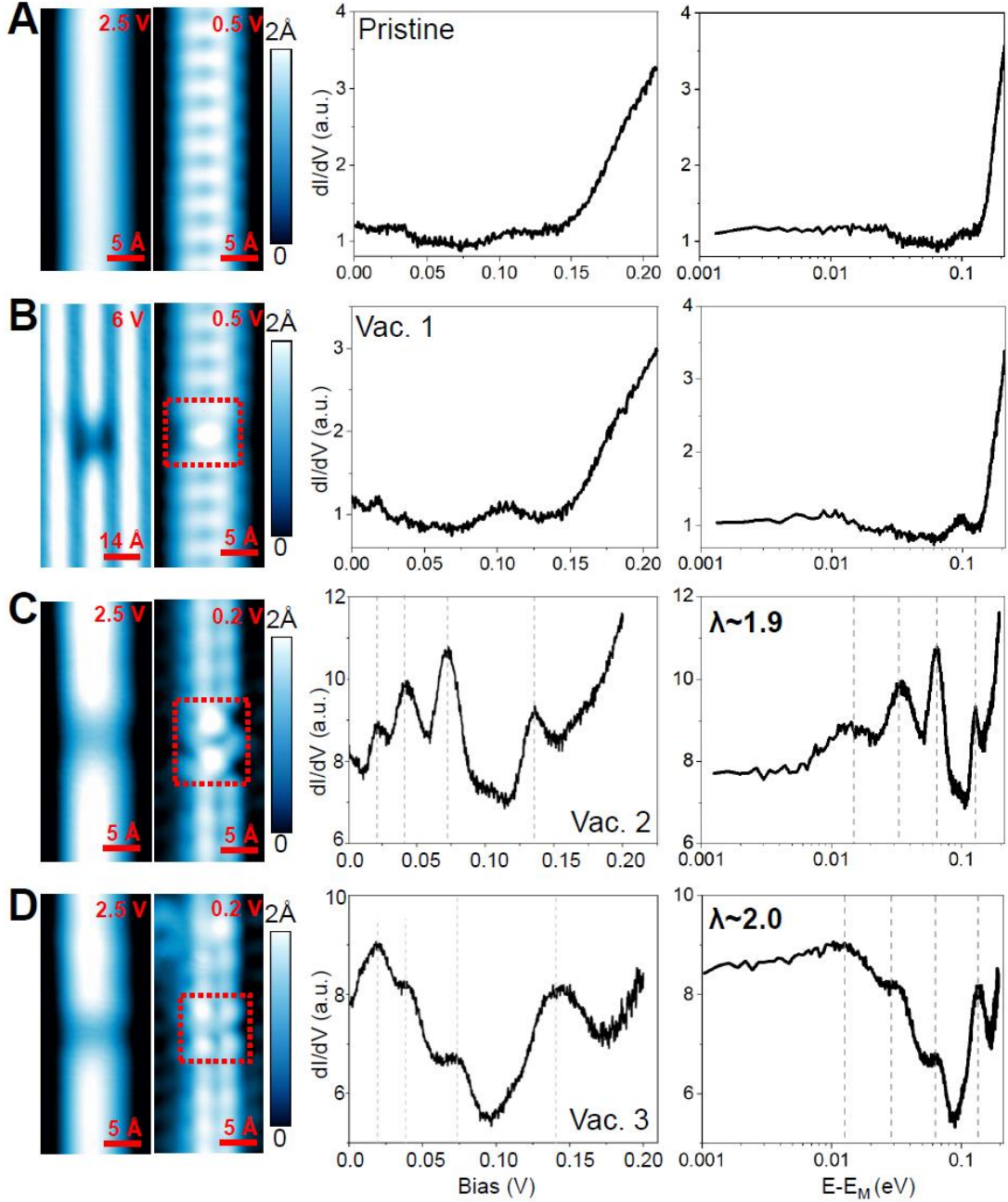

**Fig. 3. High-resolution STM images and the corresponding d$I$/d$V$ spectra plotted in both linear and log scales**, for the pristine HfTe$_5$ surface (*A*) vac. 1, single *Te$_t$* vacancy (*B*), vac. 2, a *2Hf+1Te$_d$* vacancy (*C*)**,** and vac. 3, a *Hf* vacancy (*D*)**.** a.u., arbitrary unit. $E_M$ is the middle position of the gap (~ 8 meV above $E_F$). The d$I$/d$V$ spectra shown here were measured at 4.2 K with the parameters of $V_B$ = 300 mV, $I_T$ = 300 pA and the magnitude of the bias modulation for the lock-in technique is 2.5 mV. All data were measured over 10 times in order to check the reproducibility.

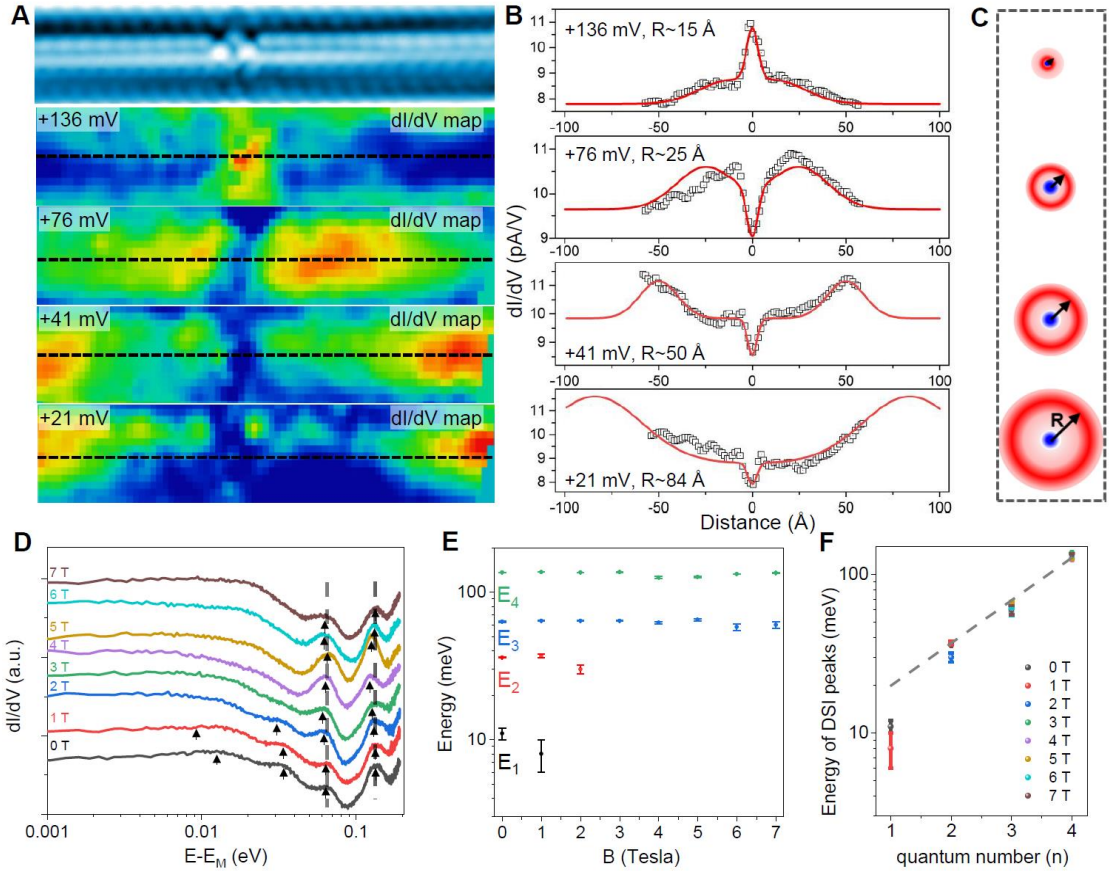

**Fig. 4. Spatial distribution and magnetic field-dependence of the quasi-bound states.** (*A*) Topographic image (top) and four d*I*/d*V* maps taken at a series of resonance energies of +21, +41, +76 and +136 mV for vac. 2. (*B*) The peak intensity of each quasi-bound state drawing from the d*I*/d*V* maps is plotted with lateral distances (black dots). The red curves are the fitting with the radius $R_n$ (the distance from the center of the vacancy) of each quasi-bound state. (*C*) Semiclassical picture of the orbits of the supercritical states. The radii exhibit the discrete scale invariance of these states. (*D*) d*I*/d*V* curves for vac. 3 at the magnetic field from 0 to 7 T. The energies of quasi-bound states are indicated by black arrows, which are identified by the Gaussian peak fitting (*SI Appendix*, Fig. S5). The grey dashed lines are guides to the eye. (*E*) The energies of quasi-bound states versus the magnetic fields. The peak positions and error bars of the quasi-bound states are determined by the Gaussian peak fitting of the d*I*/d*V* curves at different magnetic fields. (*F*) The energies of the quasi-bound states plotted on the log scale versus the index. The grey dashed line is highlighting the DSI property at 0 T.

# Supplementary Information for

Discrete scale invariance of the quasi-bound states at atomic vacancies in a topological material

Zhibin Shao[a,1], Shaojian Li[b,1], Yanzhao Liu[c,1], Zi Li[d,1], Huichao Wang[e], Qi Bian[b], Jiaqiang Yan[f], David Mandrus[f,g], Haiwen Liu[h], Ping Zhang[d,i,2], X.C. Xie[c,j,k,l], Jian Wang[c,j,k,l,2] and Minghu Pan[a,b,2]

**Corresponding authors:** Jian Wang, Minghu Pan, Ping Zhang

**Email:** jianwangphysics@pku.edu.cn (Jian Wang), minghupan@snnu.edu.cn (Minghu Pan), zhang_ping@iapcm.ac.cn (Ping Zhang)

**This PDF file includes:**

>Supplementary text
>
>Figs. S1 to S7
>
>Tables S1 to S2
>
>SI References

**Supplementary Information Text**

**Discrete scale invariant quasi-bound states at vac. 4**

In a HfTe$_5$ sample with an energy gap of about 13 mV (Fig. S2*B*), a new type of vacancy (vac. 4) is detected, as shown in Figs. S2*A* and S2*C*. By performing DFT calculations (Fig. S2*D* and Table S2), a structural model of 1Hf+2Te$_t$ vacancy is proposed and the calculations show the charge $q$ = -0.649. As shown in Fig. S2*E*, the d*I*/d*V* spectrum measured at vac. 4 shows a series of peaks at +8.8, +32, +115, and +382 mV. The weak structure at around 58 mV may result from the contribution of other angular momentum channels or the influence from other defects. By choosing the middle position of the gap ($E_M$) as the zero point, the d*I*/d*V* spectrum of vac. 4 is plotted in Fig. S2*F*, showing clear log-periodic quasi-bound states and an energy ratio λ ($E_{n+1}/E_n$) ~ 3.0. For the ratio λ ~ 3.0, an effective charge $Z$ ~ 0.62 can be estimated, which is consistent with the calculated charge value (0.649) and lower than $Z$ ~ 1 at vac. 2 and vac. 3. Thus, the discrete scale invariant quasi-bound states at vacancies with different charges in different samples strongly support the emergence of geometric quasi-bound states at the atomic scale.

**The details of fitting used in Fig. 4*B***

We take the d*I*/d*V* data exactly across the vacancy along *a* axis and fit it by including a dip function for the negative charge center and two symmetric Gaussian peak functions for the quasi-bound states. Two peak functions are symmetrically placed around the dip and the distance between the peak and the dip represents the radius of the quasi-bound state. The fitting function is:

$$I = I_0 + Ae^{-\frac{(x-R)^2}{2w^2}} + Ae^{-\frac{(x+R)^2}{2w^2}} - Be^{-\frac{x^2}{2w_1^2}}$$

where $A$ is the peak intensity and $B$ is the dip depth. $w$ and $w_1$ are the widths of the peak and the dip, respectively. $R$ is the radius of the quasi-bound state.

**Evolution of the quasi-bound states under magnetic fields**

As proposed in previous literature (1,2), the magnetic field can break the quasi-bound states by introducing a long-distance scale $l_B$, where $l_B$ is the magnetic length ($l_B = \sqrt{\frac{\hbar c}{eB}}$). When the radius of the quasi-bound state is longer than $l_B$ with increasing magnetic fields, the supercritical state with DSI would be converted into a subcritical one. During the transition, the energy of the quasi-bound state gradually approaches the Fermi energy and the corresponding resonance peak becomes wider, as shown in the following Fig. S4.

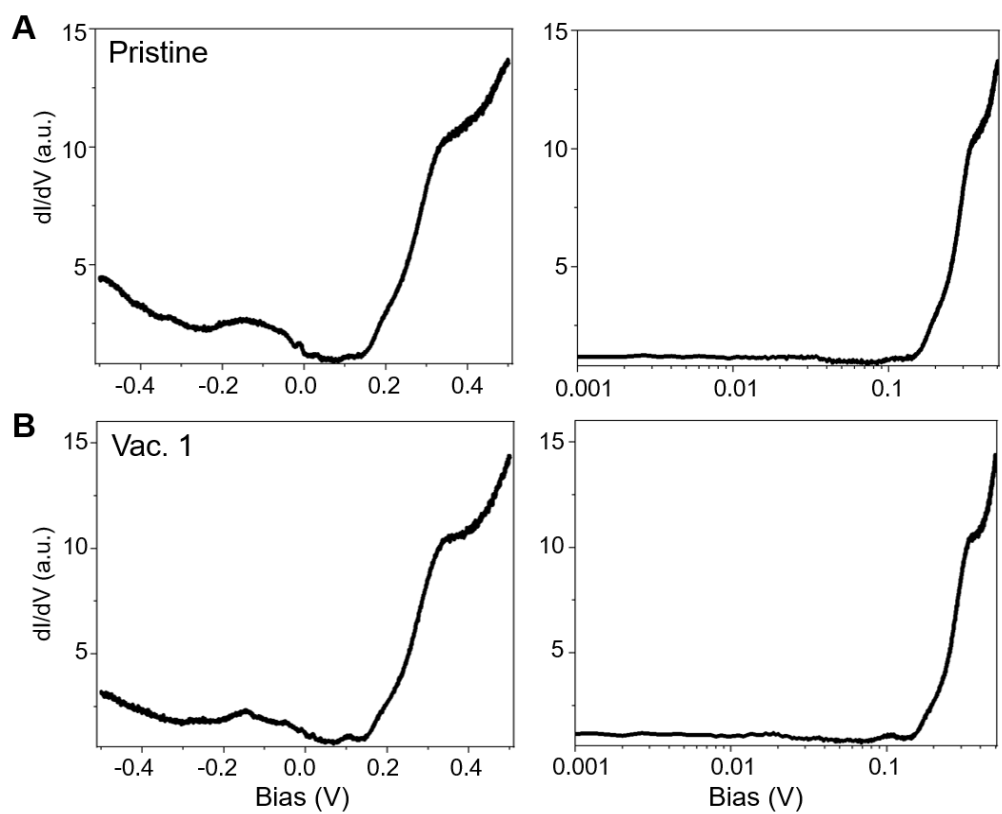

**Fig. S1.** The large energy scaled spectra of pristine surface (*A*) and vac. 1 (*B*), plotted with linear scale (left) and log scale (right).

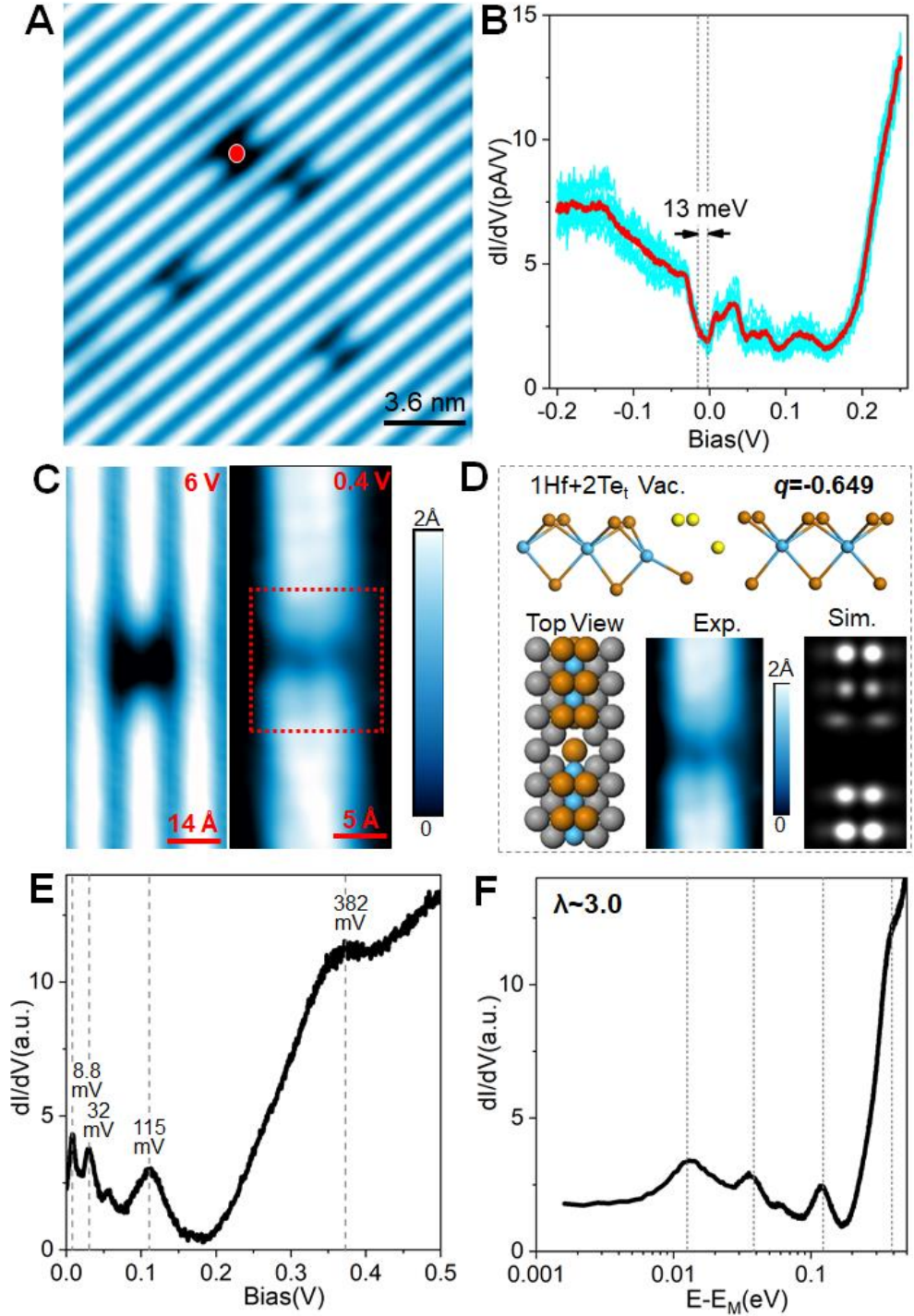

**Fig. S2.** Discrete scale invariant quasi-bound states at vac. 4. (*A*) The topographic STM image where vac. 4 is marked by the red dot. The image size is 18×18 nm$^2$ and taken with $V_B$ = 6 V. (*B*) Differential conductance d$I$/d$V$ spectra measured on the surface, which shows a small energy gap ~ 13 mV. The cyan colored curves represent eleven d$I$/d$V$ spectra and the averaged spectrum is shown as the red curve. (*C*) High-resolution STM images taken at 6 V and 0.4 V for vac. 4. (*D*) Proposed structure, experimental / simulated STM images and the calculated charge for vac. 4, assigned as $1Hf+2Te_t$ vacancy with the charge $q$ = -0.649. For the sideview of the structure (the upper panel), only the $HfTe_3$ chain containing the vacancy is shown; the vacant atoms are labeled as yellow balls. For the top view, the Te atoms in the $HfTe_3$ chain and the zigzag Te chain are labeled as brown and gray balls, respectively. (*E*) d$I$/d$V$ spectrum measured at vac. 4 and plotted on the linear scale. (*F*) d$I$/d$V$ spectrum of vac. 4 as a function of $E-E_M$ plotted on the log scale. $E_M$ is the middle position of the gap. The d$I$/d$V$ spectra shown here were measured at 4.2 K with the parameters of $V_B$ = 300 mV, $I_T$ = 300 pA and the magnitude of the bias modulation for the lock-in technique is 2.5 mV. All data were measured over 10 times in order to check the reproducibility.

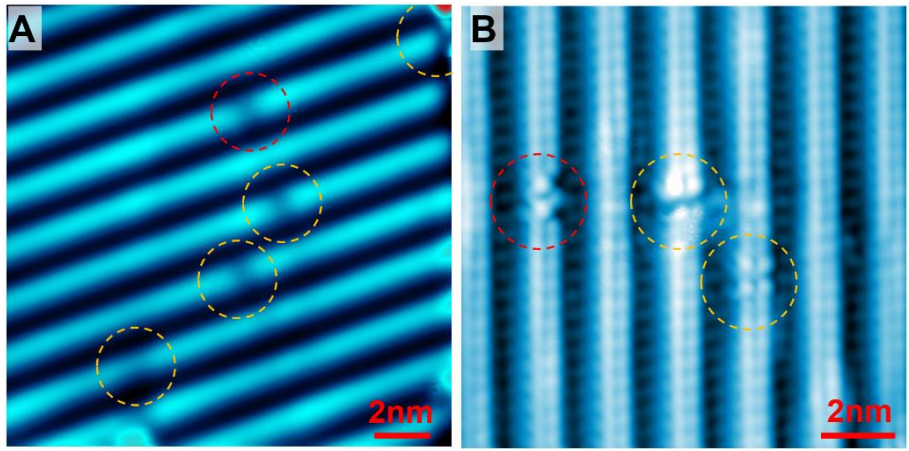

**Fig. S3.** The topographic images where vac. 2 is marked by the red dashed circles and other vacancies are marked by yellow dashed circles. The image sizes are 16×16 nm$^2$ (*A*) and 12×12 nm$^2$ (*B*). The set points are $V_B$=2.5 V, $I$ = 50 pA (*A*) and $V_B$=0.2 V, $I$ = 100 pA (*B*).

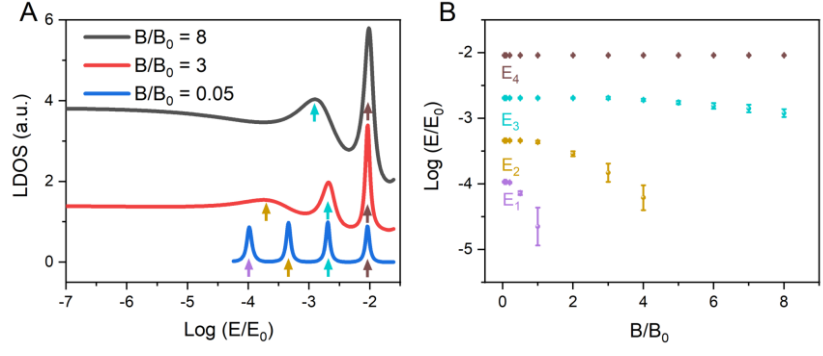

**Fig. S4.** Simulated local density of states (LDOS) for the *m*=1 angular momentum channel and effective fine structure constant *Zα* = 5.0 (obtained from the experiments on HfTe$_5$ and corresponding to λ= $E_n/E_{n-1}$ ~ 2.0) at different magnetic fields, which demonstrates the DSI transition. (*A*) Calculated LDOS as a function of log ($E/E_0$) at three selected magnetic field parameters *B*/$B_0$. Due to the DSI property of the system, the LDOS is plotted versus the dimensionless energy $E/E_0$ and $B/B_0$. The energy unit $E_0$ is set as $E_0 = \hbar v_F Z\alpha/r_0$, where $v_F$ is the Fermi velocity, $r_0$ is the length unit and can be chosen as comparable to the lattice constant. The magnetic field unit $B_0$ is set by the corresponding magnetic length $l_0 \equiv \sqrt{\frac{\hbar c}{eB_0}} = 25r_0$ (arbitrarily chosen value of $l_0$ does not change the qualitative feature of the LDOS evolution). The energies of quasi-bound states are indicated by arrows. With enlarging the magnetic field, the mean value of the low-energy quasi-bound state shifts to the Fermi energy (log ($E/E_0$) ~ $-\infty$), and meantime the corresponding resonance peak becomes broadening. (*B*) Simulated magnetic-field-dependent energies of the quasi-bound states. The peak positions and error bars of the quasi-bound states are determined by Gaussian peak fitting of the simulated LDOS curves in (*A*) at different magnetic fields. All the fitting parameters are set free during the fitting process.

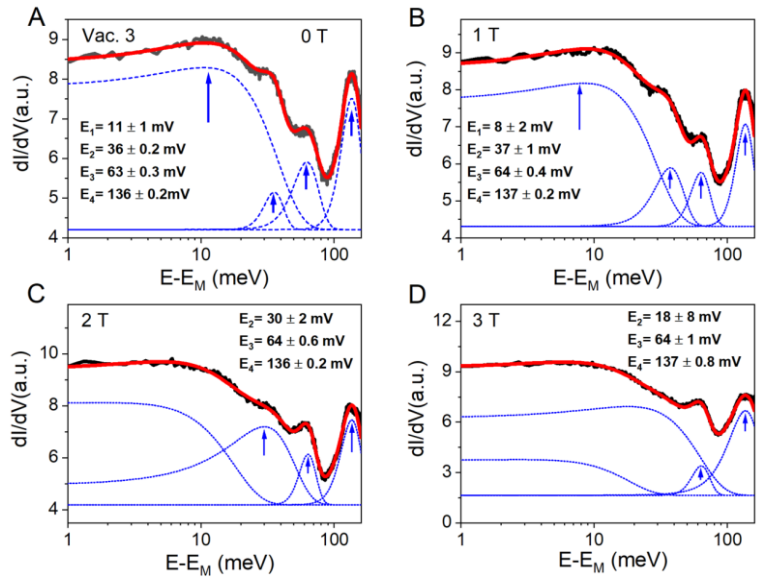

**Fig. S5.** d*I*/d*V* curves for vac. 3 at 0 T (A), 1 T (B), 2 T (C) and 3 T (D). Black lines are the experimental results and red lines are the Gaussian peak fitting curves. Blue dashed lines are the Gaussian peaks with the peak positions marked by blue arrows.

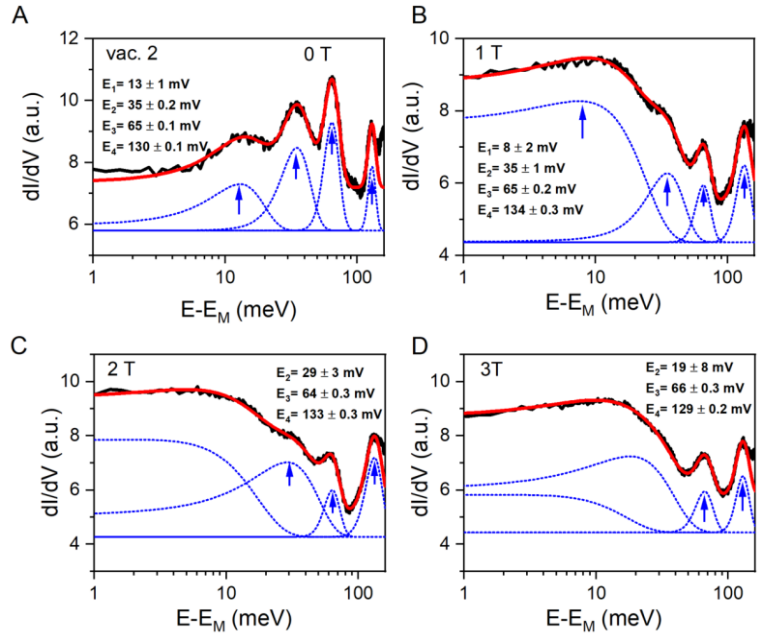

**Fig. S6.** d$I$/d$V$ curves for vac. 2 at 0 T (A), 1 T (B), 2 T (C) and 3 T (D). Black lines are the experimental results and red lines are the Gaussian peak fitting curves. Blue dashed lines are the Gaussian peaks with the peak positions marked by blue arrows.

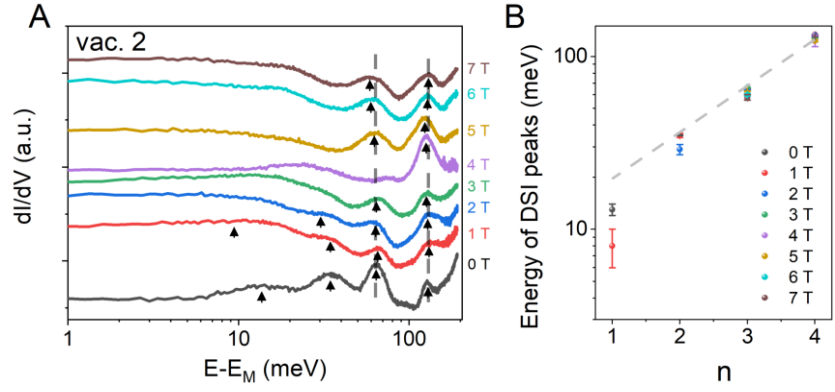

**Fig. S7.** Evolution of the d$I$/d$V$ curves with applied magnetic fields for vac. 2. (A) d$I$/d$V$ curves for vac. 2 at the magnetic field from 0 to 7 T. The energies of quasi-bound states are indicated by black arrows. The grey dashed lines are guides to the eye. (B) The energies of the quasi-bound states plotted on the log scale versus the index. The grey dashed line is highlighting the DSI property at 0 T. The peak positions and error bars of the quasi-bound states are determined by the Gaussian peak fitting of the d$I$/d$V$ curves at different magnetic fields.

**Table S1.** Comparison of experimental and theoretical defect structures, STM images and the defect charges. The most possible defects theoretically predicted are labeled by check marks in the column of Defects/Cal. Charge.

| Defects/Exp. Charge/ Exp. Image | Possible Models (Side View) | Possible Models (Top View) | Defects /Cal. Charge | Simulated Images |
|---|---|---|---|---|
| Pristine / 0 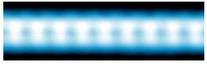 | 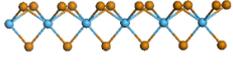 | 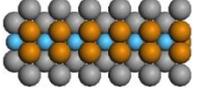 | Pristine / 0 | 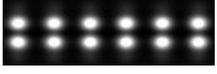 |
| Defect 1/ 0 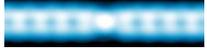 | 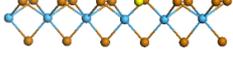 | 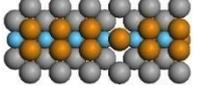 | $1Te_d$/ $-0.165$√ | 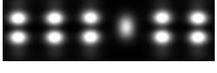 |
| Defect 2/ -1 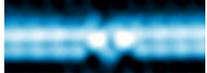 | 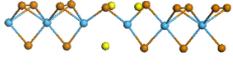 | 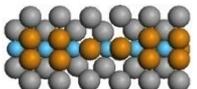 | $2Te_t+1Te_d$/ $-0.294$ | 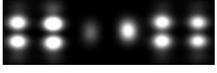 |
| | 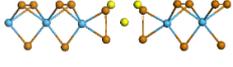 | 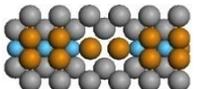 | $1Hf+2Te_t$/ $-0.595$ | 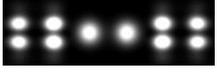 |
| | 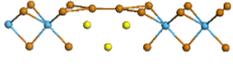 | 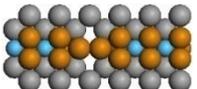 | $2Hf+1Te_d$/ $-0.972$√ | 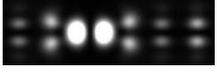 |
| Defect 3/-1 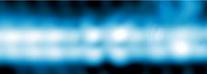 | 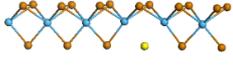 | 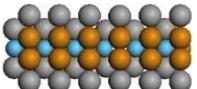 | $1Te_d$/ $-0.189$ | 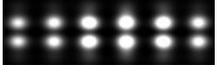 |
| | 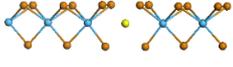 | 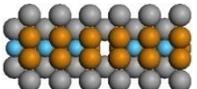 | $1Hf$/ $-0.776$√ | 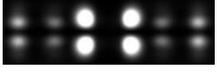 |
| | 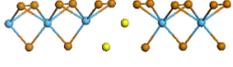 | 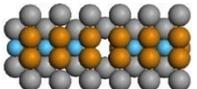 | $1Hf+1Te_d$/ $-0.725$ | 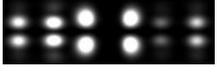 |
| | 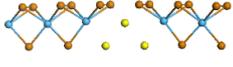 | 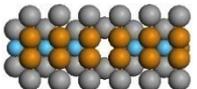 | $1Hf+2Te_d$/ $-0.558$ | 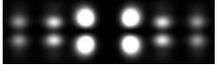 |
| | 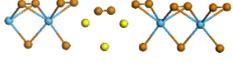 | 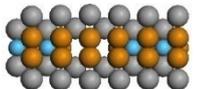 | $2Hf+1Te_d$/ $-1.009$ | 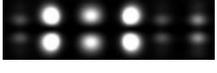 |

**Table S2.** Experimental / simulated STM images, theoretical structures, and the calculated charges for vac. 4. The most possible defect theoretically predicted is labeled by the check mark in the column of Defects/Cal. Charge.

| Defect/Exp. Charge/ Exp. Image | Possible Models (Side View) | Possible Models (Top View) | Defects /Cal. Charge | Simulated Images |
|---|---|---|---|---|
| Defect 4/-0.62 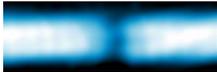 | 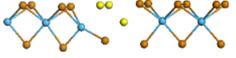 | 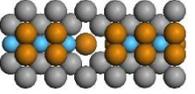 | 1Hf+2Te$_t$ / -0.649 ✓ | 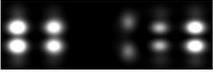 |
| | 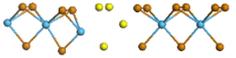 | 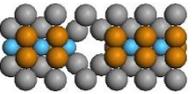 | 1Hf+2Te$_t$+1Te$_d$ /-0.428 | 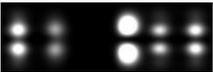 |
| | 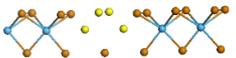 | 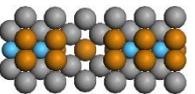 | 2Hf+2Te$_t$ / -0.557 | 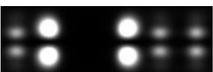 |